 \documentclass[fleqn,twoside]{article}
 \usepackage{espcrc2}

 \usepackage{graphicx}
 \usepackage[figuresright]{rotating}

 \newcommand{\AmS}{{\protect\the\textfont2
   A\kern-.1667em\lower.5ex\hbox{M}\kern-.125emS}}

 \title{XMM-Newton observation of the Chandra Deep Field-South:
     Statistical treatment of faint source spectra}

 \author{A. Streblyanska\address[MPE]{Max-Planck-Institut f\"ur
 Extraterrestrische Physik, \\
   Giessenbach-Strasse Postfach 1312, D-85741  Garching, Germany},
 J. Bergeron\address{Institut d'Astrophysique de Paris, \\ CNRS, 98 bis
 Boulevard Arago, F-75014 Paris, France},
 H. Brunner\addressmark[MPE],  A. Finoguenov\addressmark[MPE], 
 G. Hasinger\addressmark[MPE] and
 V. Mainieri\addressmark[MPE]}
       
\begin{document}
\begin{abstract}

We present first results of the X-ray spectral analysis of the
 500 ksec deep survey obtained with XMM-Newton on the Chandra Deep Field South (CDFS). Statistical distributions of spectral index and intrinsic absorption are derived for a sample containing 70 sources with a count limit of 100 (flux limit in the [2-10] keV band of 8.9 $\times 10^{-16}$ erg cm$^{-2}$ s$^{-1}$), of which 44  have redshift identification.
 We observe a separation between the type-1 and the type-2 AGN in diagnostics involving different X-ray parameters.  Using the subsample with known $z$, we show that this separation between the AGN populations is a consequence of different absorption column densities. The two populations have the same
 average spectral index, $\langle\Gamma\rangle = 2 \pm 0.1$.
 We present integrated spectrum for the most distant type-2 QSO with strong X-ray absorption and a clear soft excess; we obtained the best fit for these objects with two different models: a scattering model and a double power law model. We also confirm a progressive hardening for the combined integrated spectra for faint objects which at first was noted by \cite{toz01a}. Our results shown a clear evolution of decrease of $\langle\Gamma\rangle$ with decreasing flux in the hard 2-10 keV band.  However, we detect not only a regular increase of  $\langle\Gamma\rangle$ for different subsamples of fluxes in comparison with Chandra results, but also an internal discrepancy of the values, if we fitted in the different energy bands.

\vspace{1pc}
\end{abstract}

\maketitle

\section{INTRODUCTION}
Recent deep X-ray surveys with the Chandra and XMM-Newton observatories indicate that the bulk of the cosmic X-ray background (XRB) in the range 0.1-10 keV is the result of the integrated emission of discrete unresolved sources and accretion onto supermassive black holes over cosmic time. Such deep surveys allowed to reach extremely low flux levels and to resolve 80-90\% of the XRB below 10 keV.
From optical indentifications we know that the dominant population in these surveys is a mixture of obscured and unobscured AGN with an increasing fraction of obscuration at lower flux level (e.g. \cite{toz01b}, \cite{ros02}, \cite{fio00}, \cite{bar01} and \cite{ste02}).
Now, after resolving the majority of the XRB, the most interesting  problem is to understand the physical nature of these sources, their X-ray and optical properties, relation, influence and cosmological evolution of their different parameters.

To this aim we have performed an X-ray spectral analysis of the
CDFS faint sources with the purpose to characterize and classify the X-ray properties of different AGN populations. We search for relations between different X-ray physical parameters using the full X-ray sample, with additional analysis for the known $z$ subset. 

\section{X-RAY OBSERVATIONS}

The Chandra Deep Field-South (CDFS) is a 0.1 deg$^2$ area of the sky in the southern hemisphere has been studied intensively by means of a megasecond dataset from the Chandra Observatory (\cite{gia01}, \cite{gia02}, \cite{ros02}, \cite{toz01a}, \cite{toz01b}).  Recently the same field has been observed with the XMM-Newton Observatory. The XMM-Newton dataset has some specific advantages.  In particular, the EPIC cameras have a larger field-of-view than ACIS allowing for the detection of a number of new diffuse sources just outside the Chandra field-of-view. The EPIC instruments have unprecedent high sensitivity in the hard X-ray band, and in the [5-10] keV band our dataset is comparable to the 1Ms Chandra image.  Analysis of the Lockman Hole (LH; \cite{mai02}) has also confirmed the idea that  X-ray spectroscopy of a large number of sources is very powerful with XMM-Newton. 
The observed field was centered on the sky position RA 3:32:28 and DEC -27:48:30 (J2000). The CDFS was selected for its low galactic neutral hydrogen column density $N_{\rm H} \sim 8 \times 10^{19}$ cm$^{-2}$ (\cite{ros02}), and a lack of bright foreground  stars. The  field has been observed with XMM-Newton for a total of $\sim$ 500 ksec in July 2001 and January 2002 in guaranteed observation time (PI: J.Bergeron), but due to high background conditions and flares, some data were lost and the exposure time for good quality observation added up to $\sim$ 370 ksec. The used dataset of the CDFS is the result of the coaddition of 8 individual exposures with aimpoints only a few arcsec from each other. 

A large optical spectroscopic identification program has been carried out with the ESO Very Large Telescope (VLT) and the complete results of this campaign will be presented in \cite{szo03}.

\section{X-RAY SAMPLE}
We selected from our data a  sample of 70 X-ray sources with more than 100 counts in the 0.2-10 keV energy band, of which 44 have been spectroscopicaly identified. Our full sample includes 29 Type-1 AGN (including 12 Type-1 QSOs), 10 Type-2 AGN (including 3 Type-2 QSOs), 28 unidentified sources, one normal galaxy and one star. The used classification scheme is the same as described in  \cite{szo03}. This work includes analysis of EPIC-pn data only.

The $z$ distribution of the CDFS sources peak at redshift below 1. This distribution confirm  the already remarked disagreement between the prediction from X-ray background population synthesis models (\cite{gil01}), with 
 a maximum around $z \sim 1.5$, and the recent observations. Such a discrepancy was already noted for the 100 ksec observation of the Lockman Hole  with XMM-Newton \cite{mai02a} and for the 1 Msec Chandra survey of the CDFS \cite{ros02}.

 \section{SPECTRAL ANALYSIS}
We use an automated procedure to extract individual X-ray spectra and used XSPEC (v11.2) for the spectral fitting analysis. Initially, we fit the data with a model consisting of a power law with an intrinsic absorption (wabs or zwabs, if the redshift was known) component, with an additional photoelectric absorption component (wabs) that was fixed at the Galactic column density of $6 \times 10^{19}$ cm$^{-2}$. We used a scattering model when there was an unambiguous soft X-ray excess, which was present in several of the type-2 AGN.
 
 \subsection{Photon index  and column density}
\begin{figure}[htb]
\includegraphics[width=7cm]{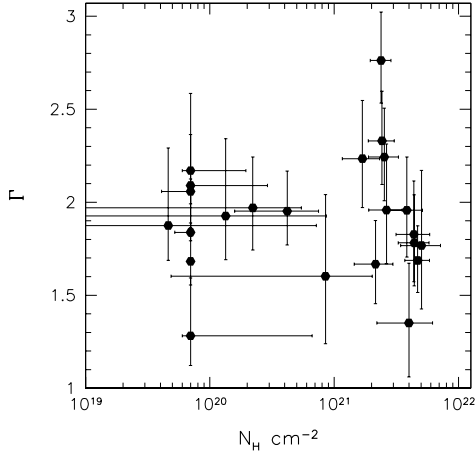} 
\caption {The observed power law photon index $\Gamma$ versus $N_{\rm H}$ for
 subsample without known $z$. For both parameters, error bars correspond to
 90\% confidence level. }
 \label{JB-E1_fig:fig1}
 \end{figure}

\begin{figure}[htb]
\includegraphics[width=7.5cm]{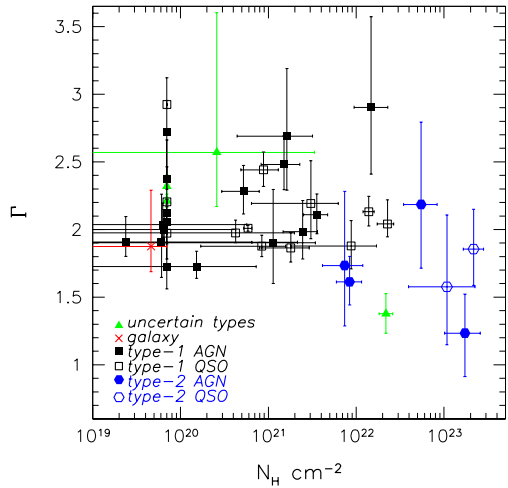} 
\caption {The rest-frame power law photon index $\Gamma$ versus $N_{\rm H}$ for the known $z$ subsample. For both parameters, error bars correspond to 90\% confidence level. }
 \label{JB-E1_fig:fig2}
 \end{figure}

From our model fits, we computed the slope of a power law spectrum (photon index $\Gamma$), an intrinsic column density $N_{\rm H}$, and the X-ray
 luminosity in the [0.5-2] and [2-10] keV rest-frame bands.  The spectral analysis was performed for the both subsample with known redshifts and the remainder of the data set.  Results are given in Figures~\ref{JB-E1_fig:fig1} and  Figures~\ref{JB-E1_fig:fig2}.

These plots show the differences between the observed values (when we assume that $z = 0$ if we do not know redshift, and that the $N_{\rm H}$ obtained is only a lower limit) and the proper values of $N_{\rm H}$ in the source rest-frame (e.g. \cite{mai02a}). 
%redshift, we assume what our z is 0 (xspec put automatically our spectra to
% z=0), and obtined the NH dereived is only a lower limit,  because while for z$>$0 your spectra is (rest-frame) redshifted to higher energy and so the real
% absorption ($N_{H}$) is higher.

The classical and obscured AGN populations show clearly different mean values of $N_{\rm H}$. We find no correlation between $\Gamma$ and the intrinsic absorption column $N_{H}$ density. 

The resulting $\Gamma$ for the type-1 AGN range from 1.73 to 2.92, with the
 majority of the sources clustering around 2.07, i.e. close to the canonical
 value for broad line unabsorbed AGNs. For type-2 AGN we obtained a lower
 value, with average  $\Gamma$ = 1.87. These results confirming the idea that the clear separation between the two AGN populations is due mostly to differences in the absorption column density, not in $\Gamma$.  
 In addition, the mean values of $\Gamma$ derived for both the known $z$ and unknown $z$ subsamples are approximately similar: we obtained $\langle\Gamma\rangle \sim 2$ with a large spread of $\pm 0.9$. 
The average value of $\Gamma$ obtained from  stacked technique is lower \cite{toz01b}  because the signature of absorption is washed out (mainly a redshift effect). 

All these results is consistent with similar analysis of the Lockman Hole
 \cite{mai02}.

 \section{INTEGRATED SPECTRA}

 \subsection{Subsamples of the sources}

\begin{figure}[htb]
\includegraphics[width=7cm]{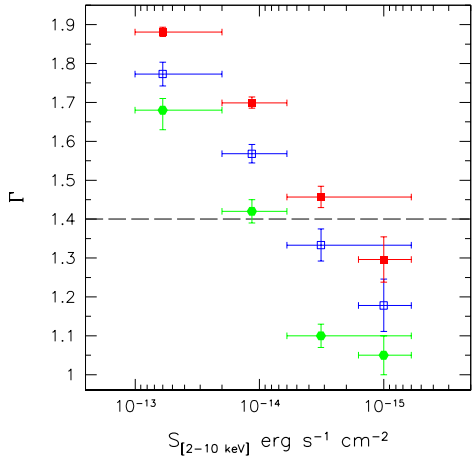}
 \caption {The average power law index  of the stacked spectra of the bright,
 medium, faint and very faint subsamples of the sources detected in the [2-10]
 keV band. The filled hexagons refer to the 1 Ms Chandra exposure of the CDFS \cite{toz01a}. XMM-Newton data were fitted in the two energy bands: 1-8 keV (open squares) for comaprison with results from the Chandra data set and 0.4-8 keV to include soft band (filled square). Errors on $\Gamma$ refer to the 90\% confidence level. The local absorption has been fixed to the Galactic value of $6 \times 10^{19}$ cm$^{-2}$. The line corresponds to the average $\Gamma$ for the background.}
 \label{JB-E1_fig:fig3}
 \end{figure}

As previously noted for the Chandra megasecond dataset  (\cite{toz01a}, \cite{toz01b} and \cite{ros02} ), in both the hard (2-10 keV) and the soft (0.5-2 keV) bands the number of hard sources increases at lower fluxes. To better understand exactly how the spectrum of XRB is built up at different fluxes, \cite{toz01a} divided the hard-band sample into four subsamplesin in terms of
 the hard fluxes: bright (S$>$ 2 $\times 10^{-14}$ erg cm$^{-2}$ s$^{-1}$), medium ( 2 $\times 10^{-14} >$ S $>$ 6 $\times 10^{-15}$ erg cm$^{-2}$ s$^{-1}$), faint (S $<$ 6 $\times 10^{-15}$ erg cm$^{-2}$ s$^{-1}$), and very faint (S $<$ 2 $\times 10^{-15}$ erg cm$^{-2}$ s$^{-1}$). Each of these samples was fit with an absorbed power law in the 1 to 10 keV energy range with the local absorption fixed to the Galactic value. \cite{toz01a} found a trends of hardening of the stacked spectra (decrease in $\langle\Gamma\rangle$) with decreasing X-ray flux in the hard  (2-10 keV) band. In order to investigate the behaviour of the spectral shape as a function of the hard flux, we repeated the same procedure that was used for the Chandra data set for the corresponding sources in our data set (assume that fluxes from XMM-Newton and Chandra datasets are equal). We use the catalog from \cite{gia02} and fitted our data in the two energy bands: 1-8 keV (for comparison with results from the Chandra dataset) and 0.4-8 keV (include soft band). For a detailed description of the data reduction and analysis see \cite{str03}.
The trend is apparent in Figure ~\ref{JB-E1_fig:fig3} where our results are compared to the corresponding results from the Chandra dataset. 

The best-fit slope of the stacked spectra for [1-8] keV is 1.773 $\pm$ 0.031,
 1.568 $\pm$ 0.024, 1.333 $\pm$ 0.041, and 1.178 $\pm$ 0.031, respectively.

The average slope of the stacked spectra for [0.4-8] keV is 1.819 $\pm$ 0.012,
 1.699 $\pm$ 0.014, 1.457 $\pm$ 0.027, and 1.296 $\pm$ 0.058, respectively.

We detect a significant hardening of the average spectral slope going to  lower fluxes. However, we also detect not only a regular increase in $\langle\Gamma\rangle$ with decreasing flux for different subsamples of fluxes in comparison with Chandra results, but we also see a discrepancy of the values within our own data set, if we fitted in the different energy bands. 

\subsection{Type-2 QSOs}
The first examples of the long-sought class of type-2 QSO have been detected in
 two deep Chandra fields \cite{nor02}, \cite{ste02} and in the
 XMM-Newton Deep survey in the Lockman Hole field \cite{leh02}.
 In our field, at $z>$2, there are six type-2 QSOs (XID 27, 54, 57, 202, 263,
 112),  the most famous of which is \#202, founded and described in detail by \cite{nor02}. All these objects have  narrow  L$_{y-\alpha}$ and CIV emission, ${HR} >$ -0.2, and faint optical magnitudes $R \ge$ 24.0 \cite{szo03}.
\begin{figure}[htb] 
\includegraphics[angle=270,width=7cm]{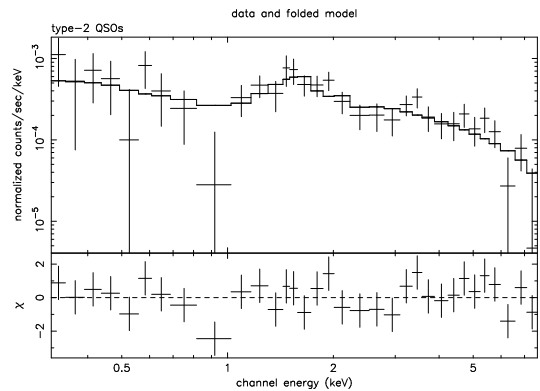}
 \caption {Integrated X-ray spectrum for the most distant type-2 QSOs.}
\label{JB-E1_fig:fig5}
 \end{figure}

\begin{figure}[htb] 
\includegraphics[angle=270,width=7cm]{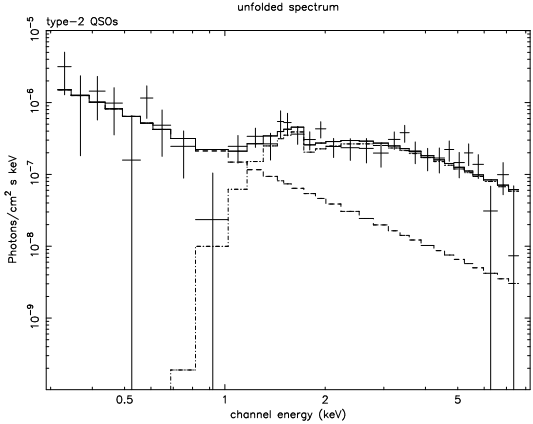}
 \caption {Integrated X-ray spectrum for the most distant type-2 QSOs; unfolded
 scattering model.}
\label{JB-E1_fig:fig6}
 \end{figure}

\begin{figure}[htb] 
\includegraphics[angle=270,width=7cm]{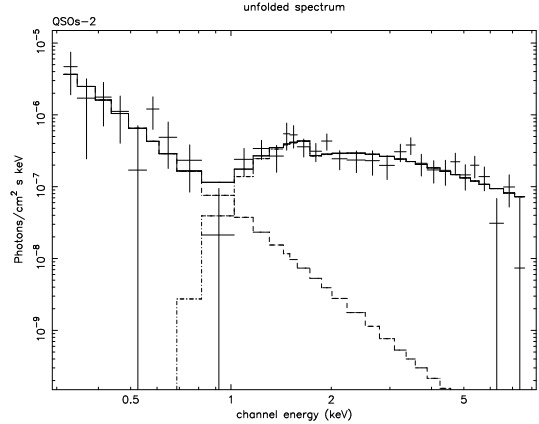}
 \caption {Integrated X-ray spectrum for the most distant type-2 QSOs; unfolded double power law model (independent two $\Gamma$ for soft and hard components).}
\label{JB-E1_fig:fig7}
\end{figure}

The next step after investigating the individual spectra was to group the faint objects by source classification and combine their spectra in order to determine general spectral properties for each type of object. For the sake of brevity, we present here only the results for the Type-$2$ QSOs. The integrated spectra for the other object classes will be presented in a future paper. 

  All of the Type-$2$ QSO spectra were extracted simultaneously, and we assumed an average redshift for these objects of $z_{int} = 3.11$ (Figures~\ref{JB-E1_fig:fig5}).

Our integrated absorbed spectrum shows a clear soft excess, which we fit by scattering model consisting of the sum of two  power law components having the same spectral index but different normalizations and absorptions. One power law is absorbed by the Galactic column density only and the other is absorbed by a high intrinsic column density (a free parameter during the fit procedure). Our model yelds a good description of the data. Result is shown in Figures~\ref{JB-E1_fig:fig6}.
 Our spectrum (1175 EPIC-pn counts in the [0.3-8] keV band) is well
 fitted ($\chi_\nu^2=0.92240$) by a scattering model  with N$_{\rm H}=8.07^{+1.47}_{-1.46}\times 10^{23}$ cm$^{-2}$, $\Gamma=2.045^{+0.069}_{-0.345}$ for the both component. It has L$_{\rm X}= 3.847\times10^{44}$ erg s$^{-1}$ in the [2-10] keV rest-frame band.

 We can test for self-consistency by fitting the spectra to more complicated spectral models. We chose to fit the spectrum to models consisting of two independent power-law components ($wabs(zwabs(powerlaw)+powerlaw)$), allowing their spectral parameters to vary independently, as the simplest way to test whether two components are required. If models in which the spectral `shapes' differ significantly are preferred, then the observed excess low-energy fluxes are more probably due to separate components of emission than to scattering. 
Results are shown in Figures~\ref{JB-E1_fig:fig7}.

By fitting a double power law model we obtain: N$_{\rm H}=5.46^{+1.82}_{-1.87}\times10^{23}$ cm$^{-2}$, and $\Gamma=1.71_{-0.37}^{+0.30}$ for the hard component and $\Gamma=3.94^{+0.61}_{-0.81}$ for the soft component ($\chi_\nu^2=0.94$). It has L$_{\rm X}= 3.652\times10^{44}$ erg s$^{-1}$ in the [2-10] keV rest-frame band.

 We can see  that our more complicated spectral models not provide significantly better fits than the  scattering model, which yelds a good description of the data. Thus, there is no reason to prefer a model with  two components for this spectrum. In general, then, it seems that the single-component and scattering models are the simplest models which adequate describe the spectra, and assumption of more complicated models is not justified by these data.

 \section{CONCLUSIONS}
In these proceedings, we present some of our results from an analysis of a $\sim 500$ ks XMM-Newton observation of the Chandra Deep Field-South. We have derived the X-ray spectral properties of a subsample of 70 sources, of which 44 have been spectroscopicaly identified. We exploited the known $z$ subsample using several X-ray diagnostics to characterise the different AGN populations.

 The differences in parameter space between the type 1 and type 2 AGN are essentially due to variations in the absorption column density.

  The average photon spectral index was found to be  $\langle\Gamma\rangle \sim 2$ for both type 1 and 2 AGN.

 Our analysis confirms the trend described by \cite{toz01b} of a
 hardening of the stacked spectra with decreasing X-ray flux in the hard (2-10
 keV) band.

Our integral spectra for the most distant type-2 QSO shows a strong X-ray
 absorption and a clear soft excess, which we  were able to fit well with a scattering model.

 \end{document}